\documentclass[12pt,reqno,a4paper]{amsart}
\usepackage{amsmath,amssymb,dsfont,graphicx,cite,latexsym,
epsf,cancel,epic,eepic}
\usepackage{amssymb,mathrsfs}
\usepackage[colorlinks=false]{hyperref}
\usepackage{color}
\usepackage{tikz}
\usepackage{caption}
\usepackage{subcaption}
\usepackage{multirow}
\usepackage{graphicx}
\usepackage{tabularx}
\usepackage{array}
\newtheorem{thm}{Theorem}[section]
\newtheorem{prop}[thm]{Proposition}
\newtheorem{cor}[thm]{Corollary}
\newtheorem{lemma}[thm]{Lemma}
\newtheorem{defn}[thm]{Definition}

\theoremstyle{definition}
\newtheorem{rk}{Remark}
\newtheorem{ex}[thm]{Example}

\newcommand{\R}{\mathbb R}

\newcommand{\Z}{\mathbb Z}
\newcommand{\N}{\mathbb N}
\newcommand{\Q}{\mathbb Q}

\newcommand{\D}{D} 

\newcommand{\T}{\overline{T}}
\newcommand{\E}{\mathrm E}
\newcommand{\V}{\mathrm V}

\newcommand{\x}{\widetilde x}
\renewcommand{\H}{\widetilde H}
\newcommand{\w}{\overline w}
\newcommand{\W}{\overline W}
\renewcommand{\v}{\gamma} 
\newcommand{\equivtau}{\bar{\tau}}
\newcommand{\equivw}{\bar{w}}
\renewcommand{\b}{\mathrm{b}}
\newcommand{\on}{\overline{\nu}}
\newcommand{\un}{\underline{\nu}}
\renewcommand{\l}{\ell}

\newcommand{\Aut}{\mathrm{Aut}}
\newcommand{\F}{\mathcal F}

\def\clap#1{\hbox to 0pt{\hss#1\hss}}

\title[]
{A combinatorial approach to integrals of Kahan-Hirota-Kimura discretizations}
\author{Ren\'e Zander}

\begin{document}

\maketitle

\begin{center}
{\footnotesize{
Institut f\"ur Mathematik, MA 7-2\\
Technische Universit\"at Berlin, Str. des 17. Juni 136,
10623 Berlin, Germany
}}
\end{center}

\begin{abstract}
\noindent
We consider an Ansatz for the study of the existence of formal integrals of motion for Kahan-Hirota-Kimura discretizations. In this context, we give a combinatorial proof of the formula of Celledoni-McLachlan-Owren-Quispel for an integral of motion of the discretization in the case of cubic Hamiltonian systems on symplectic vector spaces and Poisson vector spaces with constant Poisson structure.
\end{abstract}

\let\thefootnote\relax\footnote{E-mail: zander@math.tu-berlin.de}


\newcommand{\tree}{
\,
\begin{tikzpicture}[scale=0.15]
	\tikzstyle{vertex}=[circle,fill,scale=0.3];
	\node [circle,scale=0.3] at (0,0) {};
	\node[vertex] (r) at (0,0.5) {};
\end{tikzpicture}
\,
}

\section{Introduction}

The purpose of this paper is to explore the combinatorial structure that ensures the existence of a (formal) modified integral for the Kahan-Hirota-Kimura discrectization of cubic Hamiltonian systems on symplectic vector spaces and Poisson vector spaces with constant Poisson structure.

The Kahan-Hirota-Kimura discretization scheme has been introduced independently by Kahan \cite{Kah93}, who applied this method to the Lotka-Volterra system, and Hirota, Kimura, who applied it to the Euler top \cite{HK00E} and the Lagrange top \cite{HK00L}.
While the integrability in the Lotka-Volterra case is still an open question, the Kahan-Hirota-Kimura discretization scheme produces integrable maps for the Euler top and the Lagrange top.
Petrera, Pfadler, Suris deepened the investigation of the integrability of Kahan-Hirota-Kimura discretizations and provided an extensive list of results for concrete algebraically completely integrable systems \cite{PPS09,PS10,PPS11}.

The Kahan-Hirota-Kimura discretization scheme can be applied to any system of ordinary differential equations $\dot{x}=f(x)$ for $x:\R \to \R^n$
with
\begin{equation}
\label{vectorfield}
f(x)=Q(x)+Bx+c,\quad x\in\R^n.
\end{equation}
Here each component of $Q:\R^n \to \R^n$ is a quadratic form, while $B\in\R^{n\times n}$ and
$c \in \R^n$.
Then the Kahan-Hirota-Kimura discretization is given by
\begin{equation}
\label{KHKdiscretization}
\frac{\x-x}{2\epsilon}=Q(x,\x)+\frac12B(x+\x)+c,
\end{equation}
where 
\begin{equation*}
Q(x,\x)=\frac12\left(Q(x+\x)-Q(x)-Q(\x)\right),
\end{equation*}
is the symmetric bilinear form corresponding to the quadratic form
$Q$. Here and below we use the following notational convention
which will allow us to omit a lot of indices: for a sequence
${x}:\mathbb Z\to\mathbb R$ we write ${x}$ for ${x}_k$ and $\widetilde {x}$ for
${x}_{k+1}$. Equation (\ref{KHKdiscretization}) is {linear} with respect
to $\widetilde {x}$ and therefore defines a {rational} map
$\widetilde {x}=\Phi(x,\epsilon)$. Clearly, this map approximates the
time-$(2\epsilon)$-shift along the solutions of the original
differential system. 
(We have
chosen a slightly unusual notation $2\epsilon$ for the time step,
in order to avoid appearance of powers of 2 in numerous
formulae; a more standard choice would lead to changing
$\epsilon\mapsto\epsilon/2$ everywhere.) 
Since equation (\ref{KHKdiscretization}) remains invariant under the interchange
${x}\leftrightarrow\widetilde{{x}}$ with the simultaneous sign
inversion $\epsilon\mapsto-\epsilon$, one has the {
reversibility} property $
\Phi^{-1}(x,\epsilon)=\Phi(x,-\epsilon).
$
In particular, the map $\Phi$ is {birational}.
As already known to Kahan, the explicit form of the map $\Phi$ defined by \eqref{KHKdiscretization} is 
\begin{equation}
\x = \Phi(x,\epsilon) = x + 2\epsilon \left( I - \epsilon f'(x) \right)^{-1} f(x),\label{KHK_explicit}
\end{equation}
where $f'(x)$ denotes the Jacobi matrix of $f(x)$.

The study of the integrability of Kahan-Hirota-Kimura discretizations has been continued by
Celledoni, McLachlan, Owren, Quispel \cite{CMOQ13}.
They obtained the following remarkable result explaining some of the cases presented in \cite{PPS11}.
Let $H\colon\R^n\rightarrow\R$ be a cubic Hamiltonian, $J\in\R^{n\times n}$ be a constant skew-symmetric matrix and $f(x)=J\nabla H(x)$. Then the map (\ref{KHK_explicit}) possesses the following rational integral of motion:
\begin{equation}
\H(x,\epsilon)=H(x)+\frac{2\epsilon}{3}(\nabla H(x))^T\left(I-\epsilon f'(x)\right)^{-1}f(x).\label{CMOQ}
\end{equation}

In the field of analysis of numerical integrators there is a rich history of the use of formal series (see \cite{HLW10,CFM06,CHV05,CHV10}).
In this paper, we consider an Ansatz, proposed by Petrera and Suris, for the study of the existence of integrals of motion for the Kahan-Hirota-Kimura map (\ref{KHKdiscretization}).

\begin{defn}
A \textbf{formal integral} for the Kahan-Hirota-Kimura map (\ref{KHK_explicit}) is a formal power series
\begin{equation}
\label{formalIntegral}
\H(x,\epsilon)=\sum\limits_{q\geq 0}H_q(x)\epsilon^q,
\end{equation}
with smooth functions $H_q\colon \R^n\rightarrow\R$, $q\in\N_0$, satisfying the partial differential equations
\begin{equation}
\label{eqnEpsilon}
H^{(1)}_{q-1}[f_1]=-\sum\limits_{k=0}^{q-2}\sum\limits_{i=1}^{q-k}\sum\limits_{\substack{j_1+\dotsb+j_i=q-k\\1\leq j_1\leq\dotsb\leq j_i}}\frac{2^{i-1}}{\mu(j_1,\dotsc,j_i)}H^{(i)}_k[f_{j_1},\dotsc,f_{j_i}],
\end{equation}
where $\mu(j_1,\dotsc,j_i)=\mu_1!\mu_2!\dotsm$ and the integers $\mu_1,\mu_2,\dotsc$ count equal terms among $j_1,\dotsc,j_i$, i.e. $\mu_1,\mu_2,\dotsc$ are the multiplicities of the distinct elements $k_1,k_2,\dotsc\in\{j_1,\dotso,j_i\}$ in the tuple $(j_1,\dotsc,j_i)$.
\end{defn}

Indeed, the Kahan-Hirota-Kimura map (\ref{KHK_explicit}) has the formal series 
\begin{equation}
\x = x+2\sum\limits_{n=1}^{\infty}\epsilon^n f_n(x),\label{KHK_expansion}
\end{equation}
where $f_n(x)=(f'(x))^{n-1}f(x)$. Then substituting (\ref{KHK_expansion}) into  (\ref{formalIntegral}) and writing the Taylor series we obtain
\begin{align*}
\H(\x,\epsilon)&=\H(x,\epsilon)+\sum\limits_{i\geq1}\frac{2^i}{i!}\H^{(i)}[\epsilon f_1+\epsilon^2 f_2+\epsilon^3 f_3+\dotsb]^{i}\\
&=\H(x,\epsilon)+\sum\limits_{i\geq1}\sum\limits_{j_1,\dotsc,j_i\geq1}\frac{2^i}{i!}\epsilon^{j_1+\dotsb+j_i}\H^{(i)}[f_{j_1},\dotsc,f_{j_i}]\\
&=\H(x,\epsilon)+\sum\limits_{k\geq0}\sum\limits_{i\geq1}\sum\limits_{j_1,\dotsc,j_i\geq1}\frac{2^i}{i!}\epsilon^{k+j_1+\dotsb+j_i}H^{(i)}_k[f_{j_1},\dotsc,f_{j_i}]\\
&=\H(x,\epsilon)+\sum\limits_{q\geq1}\left(\sum\limits_{k=0}^{q-1}\sum\limits_{i=1}^{q-k}\sum\limits_{\substack{j_1+\dotsb+j_i=q-k\\j_1,\dotsc,j_i\geq1}}\frac{2^i}{i!}H^{(i)}_k[f_{j_1},\dotsc,f_{j_i}]\right)\epsilon^q\\
&=\H(x,\epsilon)+\sum\limits_{q\geq1}\left(\sum\limits_{k=0}^{q-1}\sum\limits_{i=1}^{q-k}\sum\limits_{\substack{j_1+\dotsb+j_i=q-k\\1\leq j_1\leq\dotsb\leq j_i}}\frac{2^i}{\mu(j_1,\dotsc,j_i)}H^{(i)}_k[f_{j_1},\dotsc,f_{j_i}]\right)\epsilon^q,
\end{align*}
Now, for $\H$ being a (formal) integral of motion of the Kahan-Hirota-Kimura map means that all coefficients of $\epsilon^q$ vanish for all $q\in\N$.

\begin{rk}
We put emphasis on the fact that for a formal integral $\H$ one still has to check its convergence in order to decide whether it is indeed an integral of motion.
\end{rk}

\begin{rk}
At each step $q\in\N$ the right hand side of (\ref{eqnEpsilon}) depends only on $H_0,\dotso,H_{q-2}$. Hence,
the equations (\ref{eqnEpsilon}) can be solved recursively to obtain a formal integral.
\end{rk}

\begin{ex}
We consider the equations (\ref{eqnEpsilon}) at order $\epsilon,\epsilon^2$ and $\epsilon^3$.
\begin{enumerate}
\item[$\epsilon$:]The equation reads
\begin{equation}
\label{cond1}
H^{(1)}_0[f_1]=0.
\end{equation}
Thus, $H_0$ is an integral of the continuous system.
\item[$\epsilon^2$:]The equation reads
\begin{align}
\label{cond2}
\begin{split}
H^{(1)}_1[f_1]&=-\left(H^{(1)}_0[f_2]+H^{(2)}_0[f_1,f_1]\right)\\
&=-(H^{(1)}_0[f_1])^{(1)}[f_1]=0.
\end{split}
\end{align} 
Thus, we may choose $H_1=0$.
\item[$\epsilon^3$:]The equation reads
\begin{align*}
H^{(1)}_2[f_1]=&-\left(H^{(1)}_1[f_2]
+H^{(2)}_1[f_1,f_1]\right.\\
&\left.+H^{(1)}_0[f_3]+2H^{(2)}_0[f_1,f_2]+\frac23H^{(3)}_0[f_1,f_1,f_1]\right).
\end{align*}
Using condition (\ref{cond2}) we get
\begin{equation*}
0=(H^{(1)}_1[f_1])^{(1)}[f_1]=H^{(1)}_1[f_2]+H^{(2)}_1[f_1,f_1]=0
\end{equation*}
and obtain
\begin{equation}
\label{cond3}
H^{(1)}_2[f_1]=-\left(H^{(1)}_0[f_3]+2H^{(2)}_0[f_1,f_2]+\frac23H^{(3)}_0[f_1,f_1,f_1]\right).
\end{equation}
Now, we study the existence of solutions $H_2$ to equation (\ref{cond3}). Applying condition (\ref{cond1}) we get
\begin{equation*}
0=(H^{(1)}_0[f_1])^{(1)}[f_2]=H^{(1)}_0[f_3]+H^{(1)}_0[f_1,f_2]
\end{equation*}
and equation (\ref{cond3}) can be written as
\begin{equation*}
H^{(1)}_2[f_1]=-\left(\frac23(H^{(2)}_0[f_1,f_1])^{(1)}[f_1]+\frac{1}{3}H^{(1)}_0[f_3]\right).
\end{equation*}
This means that there exists a solution $H_2$ if there is a function $\eta$ such that $\eta^{(1)}[f_1]=H^{(1)}_0[f_3]$. In the case of a Hamiltonian system on a symplectic vector space or Poisson vector space with constant Poisson structure (i.e., $f=J\nabla H_0$) by the skew-symmetry of $J$ we get 
\begin{equation*}
H^{(1)}_0[f_3]=\nabla H_0^TJH_0''JH_0''J\nabla H_0=0.
\end{equation*}
Thus, in this case we can assign $H_2=-\frac23H^{(2)}_0[f_1,f_1]$.
\end{enumerate}
\end{ex}

The purpose of the study of equations (\ref{eqnEpsilon}) is to further the understanding of the mechanism that ensures (or prevents) that the Kahan-Hirota-Kimura discretization admits integrals of motion. While experiments indicate that for the study of obstructions for the existence of formal integrals it suffices to consider (\ref{eqnEpsilon}) at low orders, to prove the existence of formal integrals one has to check that equations 
(\ref{eqnEpsilon}) are satisfied for all $q\in\N$.
A natural starting point for this investigation is given by the following claim.

\begin{prop}
\label{prop:Cel_expansion}
Let $H_0\colon\R^n\rightarrow\R$ be a cubic Hamiltonian, $J\in\R^{n\times n}$ be a constant skew-symmetric matrix and $f=J\nabla H_0$.
Define the smooth functions $H_l\colon\R^n\rightarrow\R$ by
\begin{equation}
\label{solution}
H_{2k}=\frac23(-1)^kH^{(2)}_0[f_k,f_k]\quad\text{and}\quad H_{2k-1}=0,\text{ for } k\in\N.
\end{equation}
Then, for all $q\in\N$, we have 
\begin{equation}
\label{eqnEpsilon_Cel}
\sum\limits_{k=0}^{q-1}\sum\limits_{i=0}^{q-k}\sum\limits_{j_1+\dotsb+j_i=q-k}\frac{2^i}{\mu(j_1,\dotsc,j_i)}H^{(i)}_k[f_{j_1},\dotso,f_{j_i}]=0,
\end{equation}
where $\mu(j_1,\dotsc,j_i)=\mu_1!\mu_2!\dotsm$ and $\mu_1,\mu_2,\dotsc$ count equal terms among $j_1,\dotsc,j_i$.
\end{prop}

Although the above statement is a consequence of (\ref{CMOQ}), the combinatorial structure that ensures the solvability of the partial differential equations (\ref{eqnEpsilon_Cel}) remains rather mysterious.
In this paper, we develop a combinatorial proof of this statement based on the formalism of trees described by Hairer, Lubich, Wanner \cite{HLW10}. Note that the application of this formalism is tied to the fact already noted in \cite{CMOQ13} that the solution can be expressed in terms of elementary Hamiltonians.

\section{Trees}

In this part we recall the notion of trees (related to B-series) following \cite{HLW10,CHV05,CHV10}. We consider ordinary differential equations $\dot{x}=f(x)$ with Hamiltonian vector field $f=J\nabla H$ for cubic Hamiltonians $H\colon\R^n\rightarrow\R$ and constant skew-symmetric $J\in\R^{n\times n}$.

\begin{defn}
The set 
\begin{equation*}
T=\{\text{
\begin{tikzpicture}[scale=0.15]
	\tikzstyle{vertex}=[circle,fill,scale=0.3];
	\node [circle,scale=0.3] at (0,0) {};
	\node[vertex] (r) at (0,0.5) {};
\end{tikzpicture},
\begin{tikzpicture}[scale=0.15]
	\tikzstyle{vertex}=[circle,fill,scale=0.3];
	\node[vertex] (r) at (0,0) {};
	\node[vertex] (c1) at (1,2) {};
	\draw (r) -- (c1);
\end{tikzpicture},
\begin{tikzpicture}[scale=0.15]
	\tikzstyle{vertex}=[circle,fill,scale=0.3];
	\node[vertex] (r) at (0,0) {};
	\node[vertex] (c1) at (-1,2) {};
	\node[vertex] (c2) at (1,2) {};
	\draw (r) -- (c1) (r) -- (c2);
\end{tikzpicture},
\begin{tikzpicture}[scale=0.15]
	\tikzstyle{vertex}=[circle,fill,scale=0.3];
	\node[vertex] (r) at (0,0) {};
	\node[vertex] (c1) at (1,2) {};
	\node[vertex] (c2) at (0,4) {};
	\draw (r) -- (c1) (c1) -- (c2);
\end{tikzpicture},
\begin{tikzpicture}[scale=0.15]
	\tikzstyle{vertex}=[circle,fill,scale=0.3];
	\node[vertex] (r) at (0,0) {};
	\node[vertex] (c1) at (-2,2) {};
	\node[vertex] (c2) at (2,2) {};
	\node[vertex] (c3) at (0,2) {};
	\draw (r) -- (c1) (r) -- (c2) (r) -- (c3);
\end{tikzpicture},
\begin{tikzpicture}[scale=0.15]
	\tikzstyle{vertex}=[circle,fill,scale=0.3];
	\node[vertex] (r) at (0,0) {};
	\node[vertex] (c1) at (-1,2) {};
	\node[vertex] (c2) at (1,2) {};
	\node[vertex] (c3) at (0,4) {};
	\draw (r) -- (c1) (r) -- (c2) (c1) -- (c3);
\end{tikzpicture},
\begin{tikzpicture}[scale=0.15]
	\tikzstyle{vertex}=[circle,fill,scale=0.3];
	\node[vertex] (r) at (0,0) {};
	\node[vertex] (c1) at (0,2) {};
	\node[vertex] (c2) at (-1,4) {};
	\node[vertex] (c3) at (1,4) {};
	\draw (r) -- (c1) (c1) -- (c2) (c1) -- (c3);
\end{tikzpicture},
\begin{tikzpicture}[scale=0.15]
	\tikzstyle{vertex}=[circle,fill,scale=0.3];
	\node[vertex] (r) at (0,0) {};
	\node[vertex] (c1) at (1,2) {};
	\node[vertex] (c2) at (0,4) {};
	\node[vertex] (c3) at (1,6) {};
	\draw (r) -- (c1) (c1) -- (c2) (c2) -- (c3);
\end{tikzpicture},
\dotso
}
\}
\end{equation*} 
of \textbf{rooted (unordered) trees} is recursively defined by
\begin{equation*}
\tree\in T,\qquad [\tau_1,\dotsc,\tau_m]\in T,\quad \text{ for all } \tau_1,\dotsc,\tau_m\in T,
\end{equation*}
where $\tree$ is the tree with only one vertex, and $\tau=[\tau_1,\dotsc,\tau_m]$ represents the tree obtained by grafting the roots of $\tau_1,\dotsc,\tau_m$ by additional edges to a new vertex which becomes the root of $\tau$.
The order $|\tau|$ of a tree $\tau$ is its number of vertices. A collection $\F$ of rooted trees is called \textbf{forest}.
\end{defn}

\begin{rk}
Note that $\tau=[\tau_1,\dotsc,\tau_m]$ does not depend on the ordering of $\tau_1,\dotsc,\tau_n$,
for example, $[\tree,[\tree]]=\begin{tikzpicture}[scale=0.15]
	\tikzstyle{vertex}=[circle,fill,scale=0.3];
	\node[vertex] (r) at (0,0) {};
	\node[vertex] (c1) at (-1,2) {};
	\node[vertex] (c2) at (1,2) {};
	\node[vertex] (c3) at (0,4) {};
	\draw (r) -- (c1) (r) -- (c2) (c2) -- (c3);
\end{tikzpicture}$ and $[[\tree],\tree]=\begin{tikzpicture}[scale=0.15]
	\tikzstyle{vertex}=[circle,fill,scale=0.3];
	\node[vertex] (r) at (0,0) {};
	\node[vertex] (c1) at (-1,2) {};
	\node[vertex] (c2) at (1,2) {};
	\node[vertex] (c3) at (0,4) {};
	\draw (r) -- (c1) (r) -- (c2) (c1) -- (c3);
\end{tikzpicture}$  are equal in $T$.
\end{rk}

We use the notation $\V(\tau)$ for the set of all vertices and $\E(\tau)$ for the set of all edges of $\tau\in T$. We write $e=(\nu,\nu')$ for the edge linking $\nu$ and $\nu'$. Given $\tau\in T$ we write $r(\tau)\in\V(\tau)$ for the root of $\tau$. By $\deg(\nu)$ we denote the number of edges attached to $\nu\in\V(\tau)$.

\begin{defn}
An \textbf{isomorphism} of trees $\tau_1,\tau_2$ is a map $\phi\colon\V(\tau_1)\rightarrow\V(\tau_2)$ such that $(\nu,\nu')\in\E(\tau_1)$ if and only if $(\phi(\nu),\phi(\nu'))\in\E(\tau_2)$. Trees $\tau_1,\tau_2$ are called \textbf{properly isomorphic} (we write $\tau_1=\tau_2$) and there is an isomporhism $\phi\colon\tau_1\rightarrow\tau_2$ with $\phi(r(\tau_1))=r(\tau_2)$.
\end{defn}

The set $T$ can be seen as set of equivalence classes of properly isomorphic trees.

\begin{defn}
For a tree $\tau=[\tau_1,\dotsc,\tau_m]\in T$ the \textbf{symmetry coefficient} $\sigma(\tau)$ is defined recursively by 
\begin{equation*}
\sigma(\tree)=1,\qquad \sigma(\tau)=\sigma(\tau_1)\dotsm\sigma(\tau_m)\mu_1!\mu_2!\dotsm,
\end{equation*}
where the integers $\mu_1,\mu_2,\dotsc$ count equal trees among $\tau_1,\dotsc,\tau_m$.
For a forest $\F$ the symmetry coefficient is defined by
\begin{equation*}
\sigma(\F)=\prod\limits_{\tau\in\F}\sigma(\tau).
\end{equation*}
\end{defn}

\begin{defn}
For a tree $\tau\in T$ the \textbf{branching number} $\b(\tau)$ is defined by $\b(\tau)=\deg(r(\tau))$ and the \textbf{branching factor} $\alpha(\tau)$ is defined by $\alpha(\tau)=2^{\b(\tau)}$.
\end{defn}

\begin{defn}
Let $u,v\in T$, with $u=[u_1,\dotsc,u_m]$, $v=[v_1,\dotsc,v_n]$ and $\v=(\nu_1,\dotsc,\nu_n)\in \V(u)^n$.
The \textbf{merging product} $u\times_{\v}[v_1,\dotso,v_n]$ is given by the tree obtained from $u$, where the rooted subtrees $v_1,\dotsc,v_n$ of $v$ are attached by a new edge to the vertices $\nu_1,\dotsc,\nu_n$ of $u$ respectively. 
By abuse of notation we write $u\times_{\v}v$ meaning that a representation $v=[v_1,\dotsc,v_n]$ is fixed. The \textbf{Butcher product} is defined as $u\circ v=[u_1,\dotsc,u_m,v]$.
\end{defn}

\begin{defn}
For a given smooth vector field $f\colon\D\rightarrow\R^n$ (with open $D\subset\R^n$) and for $\tau\in T$ we define the \textbf{elementary differential} $F(\tau)\colon\D\rightarrow\R^n$ by
\begin{equation*}
F(\tree)(x)=f(x),\qquad F(\tau)(x)=f^{(m)}(x)\left(F(\tau_1)(x),\dotsc,F(\tau_m)(x)\right)
\end{equation*}
for $\tau=[\tau_1,\dotsc,\tau_m]$.
\end{defn}

\begin{defn}
For a given smooth function $H\colon \D\rightarrow\R$ (with open $\D\subset\R^n$) and for $\tau\in T$ we define the \textbf{elementary Hamiltonian} $H(\tau)\colon\D\rightarrow\R$ by 
\begin{equation*}
H(\tree)(x)=H(x),\qquad H(\tau)(x)=H^{(m)}(x)\left(F(\tau_1)(x),\dotsc,F(\tau_m)(x)\right)
\end{equation*}
for $\tau=[\tau_1,\dotsc,\tau_m]$. Here, $F(\tau_i)(x)$ are elementary differentials corresponding to $f(x)=J\nabla H(x)$.
\end{defn}

Note that the solution given by (\ref{solution}) can be given in terms of elementary Hamiltonians.
The following lemma provides relations among elementary Hamiltonians that are essential for 
the validity of the relations (\ref{eqnEpsilon_Cel}).
We may stress the fact that we are in the situation of Hamiltonian systems on symplectic vector spaces or Poisson vector spaces with constant Poisson structure.

\begin{lemma}\cite{HLW10}
\label{lem:ucircv}
Elementary Hamiltonians satisfy
\begin{equation}
H(u\circ v)(x)+H(v\circ u)(x)=0\qquad \text{for all } u,v\in T.
\end{equation}
In particular, we have $H(u\circ u)(x)=0$ for all $u\in T$.
\begin{proof}
Let $u=[u_1,\dotsc,u_m]\in T$ and $v=[v_1,\dotsc,v_n]\in T$. Then using the skew-symmetry of $J$ we have 
\begin{align*}
H(u\circ v)&=H^{(m+1)}\left(F(u_1),\dotsc,F(u_m),F(v)\right)\\
&=F(v)^T(\nabla H)^{(m)}\left(F(u_1),\dotsc,F(u_m)\right)\\
&=-F(u)^T(\nabla H)^{(n)}\left(F(v_1),\dotsc,F(v_n)\right)\\
&=-H^{(n+1)}\left(F(v_1),\dotsc,F(v_m),F(u)\right)
=-H(v\circ u).
\end{align*} 
\end{proof}
\end{lemma}

\begin{rk}
\label{rk:rootchanges}
As a consequence of this lemma we note that for trees $u,v\in T$ 
which have the same graph and differ only in the position of the root, we have $H(u)(x)=(-1)^{\kappa(u,v)}H(v)(x)$, where $\kappa(u,v)$ is the number of (1 step) root changes that are necessary to obtain $u$ from $v$.
\end{rk}

\begin{defn}
Trees $\tau_1,\tau_2\in T$ are called \textbf{equivalent} (we write $\tau_1\cong\tau_2$) if they have the same graph and differ only in the position of the root (i.e. $\tau_1,\tau_2$ are isomorphic but not properly isomorhic). We denote the set of equivalence classes by $\T$.
\end{defn}

\begin{defn}
Let $H\colon \D\rightarrow\R$ (with open $\D\subset\R^n$) be a smooth function; $\tau=[\tau_1\dotsc,\tau_m]$ and $t=[t_1,\dotsc,t_n]$ be trees in $T$. Then we define the \textbf{derivative of $H(\tau)$ w.r.t $t$} as $H(\tau)[t]\colon\D\rightarrow\R$ by 
\begin{equation*}
H(\tau)[\tree](x)=H(\tau)(x),\qquad H(\tau)[t](x)=(H(\tau))^{(n)}\left(F(t_1)(x),\dotsc,F(t_n)(x)\right).
\end{equation*}
\end{defn}

\begin{lemma}
\label{lem:derivative_H}
Let $H\colon \D\rightarrow\R$ (with open $\D\subset\R^n$) be a smooth function; $\tau=[\tau_1\dotsc,\tau_m]$ and $t=[t_1,\dotsc,t_n]$ be trees in $T$. Then we have
\begin{equation}
H(\tau)[t](x)=\sum\limits_{\v=(\nu_1,\dotsc,\nu_n)\in V(\tau)^n}H(\tau\times_{\v}t)(x).
\end{equation}
\begin{proof}
This is a consequence of Leibniz' rule for derivatives.
\end{proof}
\end{lemma}

\section{Proof of proposition \ref{prop:Cel_expansion}}

In this part, we present a new combinatorial proof of proposition \ref{prop:Cel_expansion}. First, we give a reformulation of this statement using the formalism of trees. To do this in a convenient way it is necessary to introduce some further notation.

\begin{defn}
Let $T'=\left\{\tau\in T\colon \deg(\nu)\leq2\text{ for all } \nu\in\V(\tau)\text{ and } \b(\tau)=1 \right\}$ be the set of \textbf{tall trees}, i.e.,
\begin{equation*}
T'=\{\text{
\begin{tikzpicture}[scale=0.15]
	\tikzstyle{vertex}=[circle,fill,scale=0.3];
	\node [circle,scale=0.3] at (0,0) {};
	\node[vertex] (r) at (0,1) {};
\end{tikzpicture},
\begin{tikzpicture}[scale=0.15]
	\tikzstyle{vertex}=[circle,fill,scale=0.3];
	\node[vertex] (r) at (0,0) {};
	\node[vertex] (c1) at (1,2) {};
	\draw (r) -- (c1);
\end{tikzpicture},
\begin{tikzpicture}[scale=0.15]
	\tikzstyle{vertex}=[circle,fill,scale=0.3];
	\node[vertex] (r) at (0,0) {};
	\node[vertex] (c1) at (1,2) {};
	\node[vertex] (c2) at (0,4) {};
	\draw (r) -- (c1) (c1) -- (c2);
\end{tikzpicture},
\begin{tikzpicture}[scale=0.15]
	\tikzstyle{vertex}=[circle,fill,scale=0.3];
	\node[vertex] (r) at (0,0) {};
	\node[vertex] (c1) at (1,2) {};
	\node[vertex] (c2) at (0,4) {};
	\node[vertex] (c3) at (1,6) {};
	\draw (r) -- (c1) (c1) -- (c2) (c2) -- (c3);
\end{tikzpicture},
\dotso
}
\}.
\end{equation*}
\end{defn}

\begin{defn}
Let $T''=\left\{\tau=[\tau_1,\dotsc,\tau_m]\in T\colon \tau_i\in T',\text{ for } 1\leq i\leq m\right\}$ be the set of trees with branching only at the root, i.e.,
\begin{equation*}
T''=\{\text{
\begin{tikzpicture}[scale=0.15]
	\tikzstyle{vertex}=[circle,fill,scale=0.3];
	\node [circle,scale=0.3] at (0,0) {};
	\node[vertex] (r) at (0,1) {};
\end{tikzpicture},
\begin{tikzpicture}[scale=0.15]
	\tikzstyle{vertex}=[circle,fill,scale=0.3];
	\node[vertex] (r) at (0,0) {};
	\node[vertex] (c1) at (1,2) {};
	\draw (r) -- (c1);
\end{tikzpicture},
\begin{tikzpicture}[scale=0.15]
	\tikzstyle{vertex}=[circle,fill,scale=0.3];
	\node[vertex] (r) at (0,0) {};
	\node[vertex] (c1) at (-1,2) {};
	\node[vertex] (c2) at (1,2) {};
	\draw (r) -- (c1) (r) -- (c2);
\end{tikzpicture},
\begin{tikzpicture}[scale=0.15]
	\tikzstyle{vertex}=[circle,fill,scale=0.3];
	\node[vertex] (r) at (0,0) {};
	\node[vertex] (c1) at (1,2) {};
	\node[vertex] (c2) at (0,4) {};
	\draw (r) -- (c1) (c1) -- (c2);
\end{tikzpicture},
\begin{tikzpicture}[scale=0.15]
	\tikzstyle{vertex}=[circle,fill,scale=0.3];
	\node[vertex] (r) at (0,0) {};
	\node[vertex] (c1) at (-2,2) {};
	\node[vertex] (c2) at (2,2) {};
	\node[vertex] (c3) at (0,2) {};
	\draw (r) -- (c1) (r) -- (c2) (r) -- (c3);
\end{tikzpicture},
\begin{tikzpicture}[scale=0.15]
	\tikzstyle{vertex}=[circle,fill,scale=0.3];
	\node[vertex] (r) at (0,0) {};
	\node[vertex] (c1) at (-1,2) {};
	\node[vertex] (c2) at (1,2) {};
	\node[vertex] (c3) at (0,4) {};
	\draw (r) -- (c1) (r) -- (c2) (c1) -- (c3);
\end{tikzpicture},
\begin{tikzpicture}[scale=0.15]
	\tikzstyle{vertex}=[circle,fill,scale=0.3];
	\node[vertex] (r) at (0,0) {};
	\node[vertex] (c1) at (1,2) {};
	\node[vertex] (c2) at (0,4) {};
	\node[vertex] (c3) at (1,6) {};
	\draw (r) -- (c1) (c1) -- (c2) (c2) -- (c3);
\end{tikzpicture},
\begin{tikzpicture}[scale=0.15]
	\tikzstyle{vertex}=[circle,fill,scale=0.3];
	\node[vertex] (r) at (0,0) {};
	\node[vertex] (c1) at (-1,2) {};
	\node[vertex] (c2) at (1,2) {};
	\node[vertex] (c3) at (-2,4) {};
    \node[vertex] (c4) at (2,4) {};
	\draw (r) -- (c1) (r) -- (c2) (c1) -- (c3) (c2) -- (c4);
\end{tikzpicture},
\dotso
}
\}.
\end{equation*}
\end{defn}

We write $\theta_0=\tree$, $\theta_{2k}=[\tau_k,\tau_k]$, where $\tau_k$ is the tall tree with $k$ vertices, and $\theta_{2k-1}=\emptyset$ for $k\in\N$. For example:
\begin{equation*}
\theta_2=\text{
\begin{tikzpicture}[scale=0.15]
	\tikzstyle{vertex}=[circle,fill,scale=0.3];
	\node[vertex] (r) at (0,0) {};
	\node[vertex] (c1) at (-1,2) {};
	\node[vertex] (c2) at (1,2) {};
	\draw (r) -- (c1) (r) -- (c2);
\end{tikzpicture}
},\quad
\theta_4=\text{
\begin{tikzpicture}[scale=0.15]
	\tikzstyle{vertex}=[circle,fill,scale=0.3];
	\node[vertex] (r) at (0,0) {};
	\node[vertex] (c1) at (-1,2) {};
	\node[vertex] (c2) at (1,2) {};
	\node[vertex] (c3) at (-2,4) {};
    \node[vertex] (c4) at (2,4) {};
	\draw (r) -- (c1) (r) -- (c2) (c1) -- (c3) (c2) -- (c4);
\end{tikzpicture}
},\quad
\theta_6=\text{
\begin{tikzpicture}[scale=0.15]
	\tikzstyle{vertex}=[circle,fill,scale=0.3];
	\node[vertex] (r) at (0,0) {};
	\node[vertex] (c1) at (-1,2) {};
	\node[vertex] (c2) at (1,2) {};
	\node[vertex] (c3) at (-2,4) {};
    \node[vertex] (c4) at (2,4) {};
    	\node[vertex] (c5) at (-3,6) {};
    \node[vertex] (c6) at (3,6) {};
	\draw (r) -- (c1) (r) -- (c2) (c1) -- (c3) (c2) -- (c4) (c3) -- (c5) (c4) -- (c6);
\end{tikzpicture}
},\quad\dotso
\end{equation*}

We write $T_k$ for the subset of trees with $k$ vertices of the set $T$, and similarly for $T''_{k}$. 

Now, we consider the following reformulation of proposition \ref{prop:Cel_expansion}. 

\begin{prop}
\label{prop:Cel_trees}
Let $H_0\colon\R^n\rightarrow\R$ be cubic Hamiltonian, $J\in\R^{n\times n}$ be a constant skew-symmetric matrix and $f=J\nabla H_0$.
Define the smooth functions $H_l\colon\R^n\rightarrow\R$ by
\begin{equation}
\label{solution_trees}
H_{2k}=\frac23(-1)^kH_0(\theta_{2k})\quad\text{and}\quad H_{2k-1}=0,\text{ for } k\in\N.
\end{equation}
Then, for all $m\in\N$, we have 
\begin{equation}
\label{exp_eqn_trees}
\sum\limits_{k=0}^{m-1}\left(\sum\limits_{t\in T''_{m+1-k}}\frac{\alpha(t)}{\sigma(t)}H_k(t)\right)=0.
\end{equation}
\end{prop}


\begin{ex}
We consider the case $m=3$. Then equation (\ref{exp_eqn_trees}) reads
\begin{equation*}
0=
\frac23 H_0(\text{\begin{tikzpicture}[scale=0.15]
	\tikzstyle{vertex}=[circle,fill,scale=0.3];
	\node[vertex] (r) at (0,0) {};
	\node[vertex] (c1) at (-2,2) {};
	\node[vertex] (c2) at (2,2) {};
	\node[vertex] (c3) at (0,2) {};
	\draw (r) -- (c1) (r) -- (c2) (r) -- (c3);
\end{tikzpicture}})
+2H_0(\text{\begin{tikzpicture}[scale=0.15]
	\tikzstyle{vertex}=[circle,fill,scale=0.3];
	\node[vertex] (r) at (0,0) {};
	\node[vertex] (c1) at (-1,2) {};
	\node[vertex] (c2) at (1,2) {};
	\node[vertex] (c3) at (0,4) {};
	\draw (r) -- (c1) (r) -- (c2) (c1) -- (c3);
\end{tikzpicture}})
+H_0(\text{\begin{tikzpicture}[scale=0.15]
	\tikzstyle{vertex}=[circle,fill,scale=0.3];
	\node[vertex] (r) at (0,0) {};
	\node[vertex] (c1) at (1,2) {};
	\node[vertex] (c2) at (0,4) {};
	\node[vertex] (c3) at (1,6) {};
	\draw (r) -- (c1) (c1) -- (c2) (c2) -- (c3);
\end{tikzpicture}})
+H_1(\text{\begin{tikzpicture}[scale=0.15]
	\tikzstyle{vertex}=[circle,fill,scale=0.3];
	\node[vertex] (r) at (0,0) {};
	\node[vertex] (c1) at (1,2) {};
	\node[vertex] (c2) at (0,4) {};
	\draw (r) -- (c1) (c1) -- (c2);
\end{tikzpicture}})
+H_1(\text{\begin{tikzpicture}[scale=0.15]
	\tikzstyle{vertex}=[circle,fill,scale=0.3];
	\node[vertex] (r) at (0,0) {};
	\node[vertex] (c1) at (-1,2) {};
	\node[vertex] (c2) at (1,2) {};
	\draw (r) -- (c1) (r) -- (c2);
\end{tikzpicture}})
+H_2(\text{\begin{tikzpicture}[scale=0.15]
	\tikzstyle{vertex}=[circle,fill,scale=0.3];
	\node[vertex] (r) at (0,0) {};
	\node[vertex] (c1) at (1,2) {};
	\draw (r) -- (c1);
\end{tikzpicture}})
\end{equation*}
Now, we substitute $H_1$ and $H_2$ by the functions given by (\ref{solution_trees}). Then we obtain
\begin{align*}
0&=
\frac23 H_0(\text{\begin{tikzpicture}[scale=0.15]
	\tikzstyle{vertex}=[circle,fill,scale=0.3];
	\node[vertex] (r) at (0,0) {};
	\node[vertex] (c1) at (-2,2) {};
	\node[vertex] (c2) at (2,2) {};
	\node[vertex] (c3) at (0,2) {};
	\draw (r) -- (c1) (r) -- (c2) (r) -- (c3);
\end{tikzpicture}})
+2H_0(\text{\begin{tikzpicture}[scale=0.15]
	\tikzstyle{vertex}=[circle,fill,scale=0.3];
	\node[vertex] (r) at (0,0) {};
	\node[vertex] (c1) at (-1,2) {};
	\node[vertex] (c2) at (1,2) {};
	\node[vertex] (c3) at (0,4) {};
	\draw (r) -- (c1) (r) -- (c2) (c1) -- (c3);
\end{tikzpicture}})
+H_0(\text{\begin{tikzpicture}[scale=0.15]
	\tikzstyle{vertex}=[circle,fill,scale=0.3];
	\node[vertex] (r) at (0,0) {};
	\node[vertex] (c1) at (1,2) {};
	\node[vertex] (c2) at (0,4) {};
	\node[vertex] (c3) at (1,6) {};
	\draw (r) -- (c1) (c1) -- (c2) (c2) -- (c3);
\end{tikzpicture}})
-\frac23 H_0(\text{\begin{tikzpicture}[scale=0.15]
	\tikzstyle{vertex}=[circle,fill,scale=0.3];
	\node[vertex] (r) at (0,0) {};
	\node[vertex] (c1) at (-1,2) {};
	\node[vertex] (c2) at (1,2) {};
	\draw (r) -- (c1) (r) -- (c2);
\end{tikzpicture}})[\text{\begin{tikzpicture}[scale=0.15]
	\tikzstyle{vertex}=[circle,fill,scale=0.3];
	\node[vertex] (r) at (0,0) {};
	\node[vertex] (c1) at (1,2) {};
	\draw (r) -- (c1);
\end{tikzpicture}}]\\
&=
\frac23 H_0(\text{\begin{tikzpicture}[scale=0.15]
	\tikzstyle{vertex}=[circle,fill,scale=0.3];
	\node[vertex] (r) at (0,0) {};
	\node[vertex] (c1) at (-1,2) {};
	\node[vertex] (c2) at (1,2) {};
	\node[vertex] (c3) at (0,4) {};
	\draw (r) -- (c1) (r) -- (c2) (c1) -- (c3);
\end{tikzpicture}})
+H_0(\text{\begin{tikzpicture}[scale=0.15]
	\tikzstyle{vertex}=[circle,fill,scale=0.3];
	\node[vertex] (r) at (0,0) {};
	\node[vertex] (c1) at (1,2) {};
	\node[vertex] (c2) at (0,4) {};
	\node[vertex] (c3) at (1,6) {};
	\draw (r) -- (c1) (c1) -- (c2) (c2) -- (c3);
\end{tikzpicture}}).
\end{align*}
This is an identity since the trees in the last line are equivalent to 
$\text{\begin{tikzpicture}[scale=0.15]
	\tikzstyle{vertex}=[circle,fill,scale=0.3];
	\node[vertex] (r) at (0,0) {};
	\node[vertex] (c1) at (1,2) {};
	\draw (r) -- (c1);
\end{tikzpicture}}
\circ
\text{\begin{tikzpicture}[scale=0.15]
	\tikzstyle{vertex}=[circle,fill,scale=0.3];
	\node[vertex] (r) at (0,0) {};
	\node[vertex] (c1) at (1,2) {};
	\draw (r) -- (c1);
\end{tikzpicture}}
$. 
\end{ex}

\begin{rk}
\label{rk:cancellations}
Taking into account lemma \ref{lem:derivative_H} we immediately see that for each $m\in\N$ equation (\ref{exp_eqn_trees}) reads
\begin{equation}
\label{skech_1}
\sum\limits_{\tau\in T_{m+1}}a(\tau)H(\tau)=0,
\end{equation}
with rational coefficients $a(\tau)\in\Q$.
Now, using remark \ref{rk:rootchanges} we can write (\ref{skech_1}) as
\begin{equation}
\label{skech_2}
\sum\limits_{\equivtau\in\T_{m+1}}\left(\sum\limits_{\tilde{\tau}\in\equivtau}a(\tilde{\tau})(-1)^{\kappa(\tilde{\tau},\tau)}\right)H(\tau)=0.
\end{equation}
Now, at each step $m\in\N$ all summands for $\equivtau\in\T$ in equation (\ref{skech_2}) vanish.
This can be checked using the following observations.
\begin{enumerate}
\item In the situation of a cubic Hamiltonian $H_0\colon\R^n\rightarrow\R$ and a quadratic vector field $f=J\nabla H$ we have that $H_0(\tau)=0$ if there is a vertex in $\tau\in T$ with degree more than $3$.
\item $H_0(\tau)=0$ if $\tau$ is equivalent to $u\circ u$ for any tree $u\in T$ (lemma \ref{lem:ucircv}).
\item $H_0(\tau_1\circ\tau_2)=-H_0(\tau_2\circ\tau_1)$ for $\tau_1,\tau_2\in T$ (lemma \ref{lem:ucircv}), i.e., the values of elementary Hamiltonians of equivalent trees differ only in the sign.
\end{enumerate}
The first two observations will lead to the following definition of admissible trees. The last observation will require the development of a good way to enumerate all trees (and the corresponding coefficients in equation (\ref{exp_eqn_trees}) in an equivalence class of admissible trees. This will require the introduction of some notation in the following.
\end{rk}

\begin{defn}
A tree $\tau\in T$ is \textbf{admissible} if $\deg(\nu)\leq3$ for all $\nu\in\V(\tau)$ and $\tau$ is not equivalent to $\tau'\circ\tau'$ for any $\tau'\in T$. We will denote the set of admissible trees by $T^*\subset T$.
\end{defn}

\begin{defn}
Given a tree $\tau\in T^*$ a \textbf{labeling} is a map $\l\colon\V(\tau)\rightarrow\N$ such that $(-1)^{\l(\nu)}=-(-1)^{\l(\nu')}$ for adjacent vertices $\nu,\nu'\in\V(\tau)$.
\end{defn}

\begin{rk}
This definition of labeling is necessary to keep track of the different signs of the elementary Hamiltonians corresponding to the trees in an equivalence class of admissible trees.
\end{rk}

\begin{defn}
Given $\tau\in T^*$ and a subtree $w\subset\tau$ by $\tau_w$ we denote the tree obtained by contracting all vertices of $w$ to one vertex which becomes the root of $\tau_w$. We define the set of \textbf{proper subtrees} of the tree $\tau$ by
\begin{equation*}
W(\tau)=\{w\subset\tau\colon\tau_w\in T''\text{ and }w\equiv\theta_{2k},\text{ for some } k\in\N_0\}.
\end{equation*}
We use the notation $w\subset\tau$ for a strict subset $w$ of $\tau$.
The trees $w\in W(\tau)$ are viewed as rooted trees $w\in T$ with the middle vertex as root. For $w\in W(\tau)$, we write $\tau^w\in\equivtau$ for the tree in the equivalence class $\equivtau$ of $\tau$ that has the middle node of $w$ as root. Subtrees $w,w'\in W(\tau)$ are called \textbf{equivalent} if and only if there is an automorphism of $\tau$ that restricts to an isomorphism of $w$ and $w'$. The set of equivalence classes is denoted by $\W(\tau)$.
\end{defn}

We use the notation $W'(\tau)=\{w\in W(\tau)\colon|w|>1\}$ and similarly for $\W'(\tau)$.

\begin{rk}
Let $\tau\in T^*$. Then it is easy to see that for equivalent subtrees $w,w'\in W(\tau)$ we have $\tau_w=\tau_{w'}$ and $\tau^w=\tau^{w'}$. So, equivalent subtrees correspond to the same $\widetilde{\tau}\in \equivtau$. Moreover, given any labeling $\l\colon\V(\tau)\rightarrow\N$ we have that $(-1)^{\l(r(w))}=(-1)^{\l(r(w'))}$. For the last observation it is important that we exclude trees $\tau$ equivalent to $\tau'\circ\tau'$ for any $\tau'\in T$.
\end{rk}

\begin{defn}
Given $\tau\in T^*$ and a subtree $w\in W(\tau)$ we count the number of ways obtaining $\tau^w$ as merging product  $w\times_{\v}\tau_w$ with 
\begin{equation*}
\omega_{\tau}(w)=|\{\v\in\V(w)^{\b(\tau_w)}\colon w\times_{\v}\tau_w=\tau^w\}|.
\end{equation*}
We note that $\omega_{\tau}(w)$ does not depend on the choice of a representative $w\in\equivw$. 
\end{defn}

\begin{ex}
We consider an admissible tree $\tau\in T^*$ and all subtrees $w_1,w_2,w_3\in W'(\tau)$.
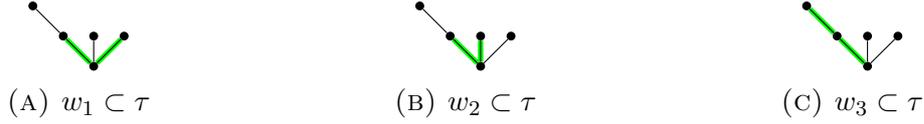
\begin{figure}[h]
\centering
\begin{subfigure}[c]{0.3\textwidth}
\centering
\begin{tikzpicture}[scale=0.2]
	\tikzstyle{vertex}=[circle,fill,scale=0.3];
	\node[vertex] (r) at (0,0) {};
	\node[vertex] (c1) at (-2,2) {};
	\node[vertex] (c2) at (2,2) {};
	\node[vertex] (c3) at (0,2) {};
	\node[vertex] (c4) at (-4,4) {};
	\draw [color=green,line width=2pt] (r) -- (c1) (r) -- (c2);
	\draw (r) -- (c1) (r) -- (c2) (r) -- (c3) (c1) -- (c4);
\end{tikzpicture}
\subcaption{$w_1\subset\tau$}
\end{subfigure}
\begin{subfigure}[c]{0.3\textwidth}
\centering
\begin{tikzpicture}[scale=0.2]
	\tikzstyle{vertex}=[circle,fill,scale=0.3];
	\node[vertex] (r) at (0,0) {};
	\node[vertex] (c1) at (-2,2) {};
	\node[vertex] (c2) at (2,2) {};
	\node[vertex] (c3) at (0,2) {};
	\node[vertex] (c4) at (-4,4) {};
	\draw [color=green,line width=2pt] (r) -- (c1) (r) -- (c3);
	\draw (r) -- (c1) (r) -- (c2) (r) -- (c3) (c1) -- (c4);
\end{tikzpicture}
\subcaption{$w_2\subset\tau$}
\end{subfigure}
\begin{subfigure}[c]{0.3\textwidth}
\centering
\begin{tikzpicture}[scale=0.2]
	\tikzstyle{vertex}=[circle,fill,scale=0.3];
	\node[vertex] (r) at (0,0) {};
	\node[vertex] (c1) at (-2,2) {};
	\node[vertex] (c2) at (2,2) {};
	\node[vertex] (c3) at (0,2) {};
	\node[vertex] (c4) at (-4,4) {};
	\draw [color=green,line width=2pt] (r) -- (c1) (c1) -- (c4);
	\draw (r) -- (c1) (r) -- (c2) (r) -- (c3) (c1) -- (c4);
\end{tikzpicture}
\subcaption{$w_3\subset\tau$}
\end{subfigure}
\end{figure}

\noindent
We notice that $w_1$ and $w_2$ are equivalent.
For $w_1$ (and $w_2$) we have
\begin{equation*}
w=
\text{\begin{tikzpicture}[scale=0.15]
	\tikzstyle{vertex}=[circle,fill,scale=0.3];
	\node[vertex] (r) at (0,0) [label={right}: \footnotesize $2$] {};
	\node[vertex] (c1) at (-1,2) [label={left}: \footnotesize $1$] {};
	\node[vertex] (c2) at (1,2) [label={right}: \footnotesize $3$] {};
	\draw (r) -- (c1) (r) -- (c2);
\end{tikzpicture}},\qquad
\tau_w=
\text{
\begin{tikzpicture}[scale=0.15]
	\tikzstyle{vertex}=[circle,fill,scale=0.3];
	\node[vertex] (r) at (0,0) {};
	\node[vertex] (c1) at (-1,2) {};
	\node[vertex] (c2) at (1,2) {};
	\draw (r) -- (c1) (r) -- (c2);
\end{tikzpicture}},\qquad
\tau^w=
\text{
\begin{tikzpicture}[scale=0.15]
	\tikzstyle{vertex}=[circle,fill,scale=0.3];
	\node[vertex] (r) at (0,0) {};
	\node[vertex] (c1) at (-2,2) {};
	\node[vertex] (c2) at (2,2) {};
	\node[vertex] (c3) at (0,2) {};
	\node[vertex] (c4) at (-4,4) {};
	\draw (r) -- (c1) (r) -- (c2) (r) -- (c3) (c1) -- (c4);
\end{tikzpicture}}.
\end{equation*}
To illustrate the computation of $\omega_{\tau}(w)$ we have labeled the vertices of $w$.
Then $\v_1=(1,2)$, $\v_2=(2,1)$, $\v_3=(2,3)$, $\v_4=(3,2)$ are all tuples of vertices such that $w\times_{\v}\tau_w=\tau^w$, i.e., $\omega_{\tau}(w)=4$.

For $w_3$ we have
\begin{equation*}
w=
\text{\begin{tikzpicture}[scale=0.15]
	\tikzstyle{vertex}=[circle,fill,scale=0.3];
	\node[vertex] (r) at (0,0) [label={right}: \footnotesize $2$] {};
	\node[vertex] (c1) at (-1,2) [label={left}: \footnotesize $1$] {};
	\node[vertex] (c2) at (1,2) [label={right}: \footnotesize $3$] {};
	\draw (r) -- (c1) (r) -- (c2);
\end{tikzpicture}},\qquad
\tau_w=
\text{
\begin{tikzpicture}[scale=0.15]
	\tikzstyle{vertex}=[circle,fill,scale=0.3];
	\node[vertex] (r) at (0,0) {};
	\node[vertex] (c1) at (-1,2) {};
	\node[vertex] (c2) at (1,2) {};
	\draw (r) -- (c1) (r) -- (c2);
\end{tikzpicture}},\qquad
\tau^w=
\text{
\begin{tikzpicture}[scale=0.15]
	\tikzstyle{vertex}=[circle,fill,scale=0.3];
	\node[vertex] (r) at (0,0) {};
	\node[vertex] (c1) at (-1,2) {};
	\node[vertex] (c2) at (1,2) {};
	\node[vertex] (c3) at (0,4) {};
	\node[vertex] (c4) at (2,4) {};
	\draw (r) -- (c1) (r) -- (c2) (c2) -- (c3) (c2) -- (c4);
\end{tikzpicture}}.
\end{equation*}
Again, we have labeled the vertices of $w$.
Then $\v_1=(1,1),\,\v_2=(3,3)$ are all tuples of vertices such that $w\times_{\v}\tau_w=\tau^w$, i.e., $\omega_{\tau}(w)=2$.
\end{ex}

\begin{defn}
Given $\tau\in T^*$ and a subtree $w\in W(\tau)$ we define 
\renewcommand{\arraystretch}{1.5}
\begin{equation*}
c(w)=\left\{ \begin{array}{ll} 
	1 & \mbox{if $|w|=1$},\\ 
	\frac23(-1)^{\frac{|w|-1}{2}} & \mbox{if $|w|>1$}.
\end{array} \right.
\end{equation*}
\end{defn}

\begin{lemma}
\label{lem:identity}
The following identity holds
\begin{equation}
\sum\limits_{k=0}^{m-1}\left(\sum\limits_{t\in T''_{m+1-k}}\frac{\alpha(t)}{\sigma(t)}H_k(t)\right)
=\sum\limits_{\equivtau\in\T^*_{m+1}}\left(\sum\limits_{\equivw\in\W}\frac{\alpha(\tau_w)}{\sigma(\tau_w)}c(w)(-1)^{\l(r(w))}\omega_{\tau}(w)\right)H(\tau).
\end{equation}
\begin{proof}
We write (\ref{exp_eqn_trees}) in a different way. Let
\begin{equation*}
c_0=1,\quad c_{2k}=\frac23(-1)^k,\quad c_{2k-1}=0,\text{ for } k\in\N.
\end{equation*}
Then using lemma \ref{lem:derivative_H} we have
\begin{align*}
\sum\limits_{k=0}^{m-1}\left(\sum\limits_{t\in T''_{m+1-k}}\frac{\alpha(t)}{\sigma(t)}H_k(t)\right)&=
\sum\limits_{k=0}^{m-1}\left(\sum\limits_{t\in T''_{m+1-k}}\frac{\alpha(t)}{\sigma(t)}c_k\sum\limits_{\v\in\V(\theta_k)^{\b(t)}}H(\theta_k\times_{\v}t)\right)\\
&=\sum\limits_{\equivtau\in\T^*_{m+1}}\left(\sum\limits_{k=0}^{m-1}\sum\limits_{t\in T''_{m+1-k}}\frac{\alpha(t)}{\sigma(t)}c_k\Omega_{t,k}(\tau)\right)H(\tau),
\end{align*}
where 
\begin{equation*}
\Omega_{t,k}(\tau)=\sum\limits_{\widetilde{\tau}\in \equivtau}(-1)^{\kappa(\tau,\widetilde{\tau})}|\{\v\in\V(\theta_k)^{\b(t)}\colon\theta_k\times_{\v}t=\widetilde{\tau}\}|.
\end{equation*}
Given $\tau\in T^*$ there is a one-to-one correspondence between triples $(t,k,\widetilde{\tau})$ and equivalence classes $\equivw$. This yields the proof.
\end{proof}
\end{lemma}

\begin{rk}
Lemma \ref{lem:identity} enhances us with a good way to enumerate and sum up the coefficients of all trees in an equivalence class of admissible trees.
In the next part we show that 
\begin{equation}
\label{formula:Cel_trees_proof}
C(\tau)=\sum\limits_{\equivw\in\W(\tau)}\frac{\alpha(\tau_w)}{\sigma(\tau_w)}c(w)(-1)^{\l(r(w))}\omega_{\tau}(w)=0
\end{equation}
for all $\tau\in T^*$. Therefor, we distinguish three cases of admissible trees $\tau\in T^*$ by the structure of their graphs:
\begin{enumerate}
\item Let $\tau\in T^*$ with $\deg(\nu)\leq2$ for all $\nu\in\V(\tau)$.
\item Let $\tau\in T^*$ with $\deg(\nu)=3$ for exactly one vertex $\nu\in\V(\tau)$.
\item Let $\tau\in T^*$ with $\deg(\nu)=3$ for at least two vertices $\nu\in\V(\tau)$.
\end{enumerate}
Then we prove equation (\ref{formula:Cel_trees_proof}) for each case.
This finishes the proof of proposition \ref{prop:Cel_trees}.
\end{rk}

\section{Proof of equation (\ref{formula:Cel_trees_proof})}

The following technical lemma will be essential for the proof of equation (\ref{formula:Cel_trees_proof}) in all three cases.

\begin{lemma}
\label{lemma:rectangle}
Let $m,n\in\N$ and $S^e_{mn}=\left\{(i,j)\in\Z^2\colon 0\leq i\leq m,\,0\leq j\leq n,\, i+j\text{ even}\right\}$ and $S^o_{mn}=\left\{(i,j)\in\Z^2\colon 0\leq i\leq m,\,0\leq j\leq n,\, i+j\text{ odd}\right\}$.
Then we have  
\begin{equation}
\sum\limits_{(i,j)\in S}(-1)^j2^{-(\delta_{i0}+\delta_{im}+\delta_{j0}+\delta_{jn})}=0,\quad S=S^e_{mn} \text{ or } S=S^o_{mn}.
\end{equation}
\begin{proof}
The claim is true for the latices $S^e_{2,2}$ and $S^o_{2,2}$ (figure (\ref{fig:S22})). Then by gluing together lattices of type $S^e_{2,2}$ or $S^o_{2,2}$ respectively we see that the claim holds for all lattices $S$ with $m,n$ even. Further we see that the claim is true for the lattices $S^e_{1,1}$ and $S^o_{1,1}$. Now, by a suitable gluing procedure we obtain that the claim holds for all lattices $S$ with $m,n\in\N$.
\begin{figure}[h]
\centering
\begin{subfigure}[c]{0.3\textwidth}
\begin{tikzpicture}[scale=1]
	\draw (-1,1) node[left] {$\frac{1}{4}$} -- (0,1) -- (1,1) node[right] {$\frac{1}{4}$};
	\draw (-1,0) -- (0,0) node[scale=0.8, below left] {$-1$} -- (1,0);
	\draw (-1,-1) node[left] {$\frac{1}{4}$} -- (0,-1) -- (1,-1) node[right] {$\frac{1}{4}$};
	\draw (-1,1) -- (-1,0) --(-1,-1);
	\draw (0,1) -- (0,0) -- (0,-1);
	\draw (1,1) -- (1,0) -- (1,-1);
\end{tikzpicture}
\end{subfigure}
\begin{subfigure}[c]{0.3\textwidth}
\begin{tikzpicture}[scale=1]
	\draw (-1,1) -- (0,1) node[above] {$\frac{1}{2}$} -- (1,1);
	\draw (-1,0) node[left] {$-\frac{1}{2}$} -- (0,0) -- (1,0) node[right] {$-\frac{1}{2}$};
	\draw (-1,-1) -- (0,-1) node[below] {$\frac{1}{2}$} -- (1,-1);
	\draw (-1,1) -- (-1,0) --(-1,-1);
	\draw (0,1) -- (0,0) -- (0,-1);
	\draw (1,1) -- (1,0) -- (1,-1);
\end{tikzpicture}
\end{subfigure}
\caption{$S^e_{2,2}$ and $S^o_{2,2}$}
\label{fig:S22}
\end{figure}
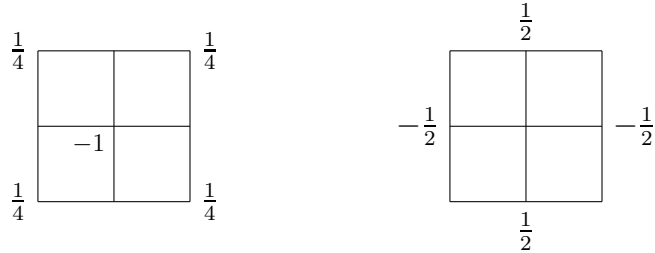
\end{proof}
\end{lemma}

\begin{cor}
\label{cor:line}
Let $m\in\N$ and $S_m=\left\{i\in\Z\colon 0\leq i\leq m\right\}$. Then we have
\begin{equation}
\sum\limits_{i\in S_m}(-1)^i2^{-(\delta_{i0}+\delta_{im})}=0.
\end{equation}
\begin{proof}
This follows from lemma \ref{lemma:rectangle} by contracting the lattice $S^e_{m,1}$ to $S_m$.
\end{proof}
\end{cor}


We distinguish three types of graphs of admissible trees $\tau\in T^*$.

\begin{enumerate}

\item Let $\tau\in T^*$ with $\deg(\nu)\leq2$ for all $\nu\in\V(\tau)$.

In this situation the cardinality $|\tau|$ is always odd (otherwise $\tau\cong\tau'\circ\tau'$ for some $\tau'\in T$). The graph of $\tau$ is illustrated in figure (\ref{fig:caseI}). Note that we label the vertices sequently by $\nu_{-a},\dotsc,\nu_{a}$. Let $w_0\subset\tau$ be the subtree consisting of the vertex $\nu_0$.

\begin{figure}[h]
\centering
\begin{tikzpicture}[scale=0.6]
	\tikzstyle{vertex}=[circle,fill,scale=0.3]
	\node [circle,scale=0.3] at (0,1) {};
	\node[vertex] (a1_1) at (-2,0) [label={below}: \footnotesize $\nu_{-1}$] {};
	\node[vertex] (a2_1) at (2,0) [label={below}: \footnotesize $\nu_1$] {};
	\node[vertex] (a1_2) at (-4,0) [label={below}: \footnotesize $\nu_{-a}$] {};
	\node[vertex] (a2_2) at (4,0) [label={below}: \footnotesize $\nu_a$] {};
	\node[vertex, color=green] (v0) at (0,0) [label={below}: \footnotesize $\nu_0$] [label={above right}: \footnotesize \textcolor{green}{$w_0$}] {};
    \draw (a1_1) -- (v0) (a2_1) -- (v0);
    \draw[dashed] (a1_2) -- (a1_1) (a2_2) -- (a2_1);
\end{tikzpicture}
\caption{Type ($1$) graph.}
\label{fig:caseI}
\end{figure}
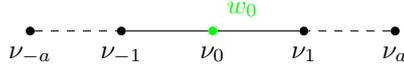

\item Let $\tau\in T^*$ with $\deg(\nu)=3$ for exactly one vertex $\nu\in\V(\tau)$.

The graph of $\tau$ is illustrated in figure (\ref{fig:caseII}).
Let $\nu_0$ denote the unique vertex with degree equal to $3$. (Then the tree $\tau\in T^*$ corresponds to the choice of $\nu_0$ as root). Let $w_0\subset\tau$ be the subtree consisting only of the vertex $\nu_0$. We denote the branches attached to $\nu_0$ by $\alpha_1,\alpha_2,\alpha_3$. We write $a_i=|\alpha_i|$, for $i=1,2,3$.

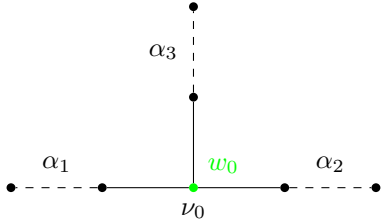
\begin{figure}[h]
\centering
\begin{tikzpicture}[scale=0.6]
	\tikzstyle{vertex}=[circle,fill,scale=0.3]
	\node[vertex] (a1_1) at (-2,0) {};
	\node[vertex] (a2_1) at (2,0) {};
	\node[vertex] (a3_1) at (0,2) {};
	\node[vertex] (a1_2) at (-4,0) {};
	\node[vertex] (a2_2) at (4,0) {};
	\node[vertex] (a3_2) at (0,4) {};
	\node[circle,scale=0.3] (n1) at (-3,0) [label={above}: \footnotesize $\alpha_1$] {};
	\node[circle,scale=0.3] (n2) at (3,0) [label={above}: \footnotesize $\alpha_2$] {};
	\node[circle,scale=0.3] (n3) at (0,3) [label={left}: \footnotesize $\alpha_3$] {};
	\node[vertex, color=green] (v0) at (0,0) [label={below}: \footnotesize $\nu_0$] [label={above right}: \footnotesize \textcolor{green}{$w_0$}] {};
	\draw (a1_1) -- (v0) (a2_1) -- (v0) (a3_1) -- (v0);
    \draw[dashed] (a1_1) -- (a1_2) (a2_1) -- (a2_2) (a3_1) -- (a3_2);
\end{tikzpicture}
\caption{Type ($2$) graph.}
\label{fig:caseII}
\end{figure}

\item Let $\tau\in T^*$ with $\deg(\nu)=3$ for at least two vertices $\nu\in\V(\tau)$.

The graph of $\tau$ is illustrated in figure (\ref{fig:caseIII}).
Let $\on,\un$ be the extremal vertices with degree equal to $3$, i.e., all other vertices with degree equal to three lie on the segment between $\on$ and $\un$. Let $w_0$ denote the subtree connecting $\on$ and $\un$. (The tree $\tau\in T^*$ corresponds to the choice of a vertex $\nu_0\subset w_0$ as root).
We denote the branches attached to $\on$ and $\un$ by $\alpha_1,\alpha_2$ and $\alpha_3,\alpha_4$ respectively. We write $a_i=|\alpha_i|$, for $i=1,\dotsc,4$.

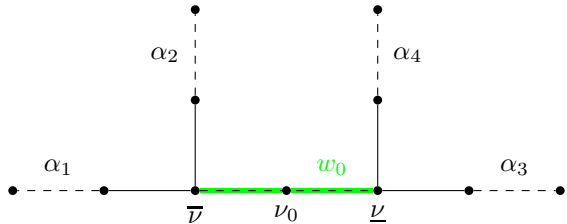
\begin{figure}[h]
\centering
\begin{tikzpicture}[scale=0.6]
	\tikzstyle{vertex}=[circle,fill,scale=0.3];
	\node[vertex] (v0) at (0,0) [label={below}: \footnotesize $\nu_0$] {};
	\node[vertex] (vs) at (-2,0) [label={below}: \footnotesize $\on$] {};
	\node[vertex] (vt) at (2,0) [label={below}: \footnotesize $\un$] {};
	\node[vertex] (a1_1) at (-4,0) {};
	\node[vertex] (a2_1) at (-2,2) {};
    \node[vertex] (b1_1) at (4,0) {};
	\node[vertex] (b2_1) at (2,2) {};
	\node[vertex] (a1_2) at (-6,0) {};
	\node[vertex] (a2_2) at (-2,4) {};
    \node[vertex] (b1_2) at (6,0) {};
	\node[vertex] (b2_2) at (2,4) {};
	\node[circle,scale=0.3] at (-2,3) [label={left}: \footnotesize $\alpha_2$] {};
	\node[circle,scale=0.3] at (2,3) [label={right}: \footnotesize $\alpha_4$] {};
	\node[circle,scale=0.3] at (-5,0) [label={above}: \footnotesize $\alpha_1$] {};
	\node[circle,scale=0.3] at (5,0) [label={above}: \footnotesize $\alpha_3$] {};
	\node[circle,scale=0.3] at (1,0) [label={above}: \footnotesize \textcolor{green}{$w_0$}] {};
	\draw[color=green,line width=2pt] (vs) -- (v0) (v0) -- (vt);
	\draw (vs) -- (a1_1) (vs) -- (a2_1) (vt) -- (b1_1) (vt) -- (b2_1);	
	\draw[dashed] (vs) --(vt) (a1_1) -- (a1_2) (a2_1) -- (a2_2) (b1_1) -- (b1_2) (b2_1) -- (b2_2);	
\end{tikzpicture}
\caption{Type ($3$) graph.}
\label{fig:caseIII}
\end{figure}

\end{enumerate}

\begin{defn}
Let $\tau\in T^*$ and $w\in W(\tau)$, with $|w|>1$. Then the pair $(\tau,w)$ is called \textbf{symmetric}
if there is an automorphism of $\phi\in\Aut(\tau)$ that restricts to an automorphism $\phi|_w\in\Aut(w)$ of $w\subset\tau$ such that $\phi|_{w}=r$, where $r\in\Aut(w)$ denotes the reflection at the middle vertex of $w$. We define the \textbf{symmetry coefficient} of the pair $(\tau,w)$ by
\renewcommand{\arraystretch}{1.5}
\begin{equation*}
\varsigma(\tau,w)=\left\{ \begin{array}{ll} 
	1 & \mbox{if $(\tau,w)$ symmetric},\\ 
	2 & \mbox{if $(\tau,w)$ non-symmetric}.
\end{array} \right.
\end{equation*}
\end{defn}

\begin{lemma}
\label{lem:proof}
Let $\tau\in T^*$ and $w\in W(\tau)$, with $|w|>1$. Then we have
\begin{align}
\omega_{\tau}(w)&=\varsigma(\tau,w)\frac{\sigma(\tau_w)}{\sigma(\tau\setminus \E(w))},\tag{i}\label{eqn:omega}\\
|\w|&=\frac{\varsigma(\tau,w)}{2}\frac{|\Aut(\tau)|}{\sigma(\tau\setminus \E(w))},\tag{ii}\label{eqn:card_w}
\end{align}
where $\tau\setminus \E(w)$ denotes the forest obtained from $\tau$ by deleting all edges contained in $w\subset\tau$.
\begin{proof} We prove both claims.

\smallskip
\noindent (i) Let $\tau_w=[\tau_1,\dotsc,\tau_N]$. 

Observe that if $\tau_{n}=\tau_{n'}$, for $1\leq n,n'\leq N$, and $\v=(\dotsc,\nu_{n},\dotsc,\nu_{n'},\dotsc)$ and $\v'=(\dotsc,\nu_{n'},\dotsc,\nu_n,\dotsc)$ is the same tuple of vertices with $\nu_{n}$ and $\nu_{n'}$ interchanged, then $w\times_{\v}\tau_w=w\times_{\v'}\tau_w$. This yields the factor $\sigma(\tau_w)$. 

Suppose that $\tau_{n}=\tau_{n'}$, for $1\leq n,n'\leq N$, and $\nu_{n}=\nu_{n'}$. Then the tuples
$\v=(\dotsc,\nu_{n},\dotsc,\nu_{n'},\dotsc)$ and $\v'=(\dotsc,\nu_{n'},\dotsc,\nu_n,\dotsc)$ are actually identical. Hence, we divide by $\sigma(\tau\setminus \E(w))$.

Suppose that the vertices of $w$ are sequently labeled by $1,\dotsc,|w|$. For $\v\in\V(w)$ we denote by $\bar{\v}$ the tuple of vertices obtained from $\v$ by replacing each $\nu\in\v$ with $|w|+1-\nu$. Obviously, we have $w\times_{\v}\tau_w=w\times_{\bar{\v}}\tau_w$. So, this yields the factor $2$ in the non-symmetric case (i.e. $\tau$ is non-symmetric or $\tau$ is symmetric and $(\tau,w)$ is non-symmetric). If $(\tau,w)$ is symmetric $\bar{\v}$ is already counted with $\sigma(\tau_w)$.

\smallskip
\noindent (ii) We divide the total number of automorphisms of $\tau$ by the number of automorphisms that restrict to an automorphism of $w\subset\tau$. Then the claim follows by the orbit-stabilizer theorem.

\end{proof}
\end{lemma}

\begin{cor}
\label{cor:proof}
Let $\tau\in T^*$ and $w\in W(\tau)$, with $|w|>1$. Then we have
\begin{equation}
\omega_{\tau}(w)|\Aut(\tau)|=2\sigma(\tau_w)|\w|.
\end{equation}
\begin{proof}
This is a direct consequence of lemma \ref{lem:proof}.
\end{proof}
\end{cor}

Now, we are in the position to complete the proof of equation (\ref{formula:Cel_trees_proof}).

\begin{proof} We proof all claims.

\smallskip
\noindent (1) Applying corollary \ref{cor:proof} and taking into account that for $|w|=1$ we have $|\w|=2/\sigma(\tau_w)$ and $\omega_{\tau}(w)=1$ we obtain
\begin{equation*}
C(\tau)
=\frac{1}{2}\sum\limits_{\substack{w\in W(\tau)\\|w|=1}}\alpha(\tau_w)c(w)(-1)^{\l(r(w))}
+\sum\limits_{\substack{w\in W(\tau)\\|w|>1}}\alpha(\tau_w)c(w)(-1)^{\l(r(w))}.
\end{equation*}

Let $w_{ij}\in W(\tau)$ denote the subtree connecting the vertices $\nu_i$ and $\nu_j$, for $i\leq j$. Then, for $w=w_{ij}$, we have:

\begin{enumerate}

\item[(i)] By definition of $c(w)$ and $\alpha(\tau_w)$ we compute
\renewcommand{\arraystretch}{1.5}
\begin{equation*}
\label{caseI_calpha}
c(w)\alpha(\tau_w)=\left\{ \begin{array}{ll} 
	\alpha(\tau_{w_0})2^{-(\delta_{-ai}+\delta_{ai})} & \mbox{if $|w|=1$},\\ 
	\dfrac23\alpha(\tau_{w_0})(-1)^{\frac{j-i}{2}}2^{-(\delta_{-ai}+\delta_{aj})} & \mbox{if $|w|>1$}.
\end{array} \right.
\end{equation*}
Note that if $w$ ends at $\nu_{-a}$ (i.e. $i=-a$) the branching number $\b(\tau_w)$ is reduced by $1$ compared to $\b(\tau_{w_0})$. The same holds if $w$ ends at $\nu_{a}$ (i.e., $j=a$).

\item[(ii)] We can assume that $(-1)^{\l(r(w))}=(-1)^{\frac{i+j}{2}}$.

\end{enumerate}

Finally, we have
\begin{equation}
\label{caseI_proof_1}
\sum\limits_{\substack{w\in W(\tau)\\|w|=1}}\alpha(\tau_w)c(w)(-1)^{\l(r(w))}
=\alpha(\tau_{w_0})\sum\limits_{\substack{w\in W(\tau)\\|w|=1}}(-1)^i2^{-(\delta_{-ai}+\delta_{ai})},
\end{equation}
and
\begin{equation}
\label{caseI_proof_2}
\sum\limits_{\substack{w\in W(\tau)\\|w|>1}}\alpha(\tau_w)c(w)(-1)^{\l(r(w))}
=\frac{2\alpha(\tau_{w_0})}{3}\sum\limits_{\substack{w\in W(\tau)\\|w|>1}}(-1)^j2^{-(\delta_{-ai}+\delta_{aj})}.
\end{equation}
Using corollary \ref{cor:line} we see that (\ref{caseI_proof_1}) and (\ref{caseI_proof_2}) vanish. Note that in (\ref{caseI_proof_2}) the sum vanishes along every diagonal with $j-i=c$, for $c\in\N$.

\smallskip
\noindent (2) Applying corollary \ref{cor:proof} and taking into account that $|\w_0|=\omega_{\tau}(w_0)=1$ we obtain
\begin{equation*}
C(\tau)=\frac{\alpha(\tau_{w_0})c(w_0)(-1)^{\l(r(w_0))}}{|\Aut(\tau)|}+\frac{2}{|\Aut(\tau)|}\sum\limits_{w\in W'(\tau)}\alpha(\tau_w)c(w)(-1)^{\l(r(w))}.
\end{equation*}

Let $w_{ij}^{kl}\in W(\tau)$ denote the subtree connecting the $i$th vertex of the branch $\alpha_k$ and the $j$th vertex of the branch $\alpha_l$, for $k,l=1,2,3$, and $W_{kl}=\{w_{ij}^{kl}\in W(\tau)\}$. Then, for $w=w_{ij}^{kl}$, we have:

\begin{enumerate}

\item[(i)] By definition of $c(w)$ and $\alpha(\tau_w)$ we compute
\renewcommand{\arraystretch}{1.5}
\begin{equation*}
c(w)\alpha(\tau_w)=\left\{ \begin{array}{ll}  
	\alpha(\tau_{w_0}) & \mbox{if $|w|=1$},\\ 
	\dfrac23\alpha(\tau_{w_0})(-1)^{\frac{i+j}{2}}2^{-(\delta_{ia_k}+\delta_{ja_l})} & \mbox{if $|w|>1$}.
\end{array} \right.
\end{equation*}
Note that if $w$ ends at the $a_k$th vertex of $\alpha_k$ (i.e., $i=a_k$) the branching number $\b(\tau_w)$ is reduced by $1$ compared to $\b(\tau_{w_0})$. The same holds if $w$ ends at $a_l$th vertex of $\alpha_l$ (i.e., $j=a_l$).

\item[(ii)] We can assume that $(-1)^{\l(r(w))}=(-1)^{\frac{j-i}{2}}$.

\item[(iii)] Passing from summation over all $w\in W'(\tau)$ to summation over all $w\in W_{kl}'$, for $1\leq k<l\leq 3$, we divide by $2^{\delta_{i0}+\delta_{j0}}$ since the affected subtrees (with $i=0$ or $j=0$) are counted multiple.

\end{enumerate}

Finally, we obtain
\begin{equation*}
C(\tau)=\frac{4\alpha(\tau_{w_0})}{3|\Aut(\tau)|}\sum\limits_{1\leq k<l\leq 3}\left(\frac14+\sum\limits_{w\in W_{kl}'}(-1)^j2^{-(\delta_{ia_k}+\delta_{ja_l}+\delta_{i0}+\delta_{j0})}\right).
\end{equation*}
Now, the claim follows by lemma \ref{lemma:rectangle}.

\smallskip
\noindent (3) Applying corollary \ref{cor:proof} we obtain
\begin{equation*}
C(\tau)=\frac{2}{|\Aut(\tau)|}\sum\limits_{w\in W(\tau)}\alpha(\tau_w)c(w)(-1)^{\l(r(w))}.
\end{equation*}

Let $w_{ij}^{kl}\in W(\tau)$ denote the subtree connecting the $i$th vertex of the branch $\alpha_k$ and the $j$th vertex of the branch $\alpha_l$, for $k=1,2$, $l=3,4$, and $W_{kl}=\{w_{ij}^{kl}\in W(\tau)\}$. Then, for $w=w_{ij}^{kl}$, we have:

\begin{enumerate}

\item[(i)] By definition of $c(w)$ and $\alpha(\tau_w)$ we compute
\begin{equation*}
c(w)\alpha(\tau_w)=\frac23\alpha(\tau_{w_0})(-1)^{\frac{i+j+|w_0|-1}{2}}2^{-(\delta_{ia_k}+\delta_{ja_l})}.
\end{equation*}
Note that if $w$ ends at the $a_k$th vertex of $\alpha_k$ (i.e., $i=a_k$) the branching number $\b(\tau_w)$ is reduced by $1$ compared to $\b(\tau_{w_0})$. The same holds if $w$ ends at $a_l$th vertex of $\alpha_l$ (i.e., $j=a_l$).

\item[(ii)] We can assume that $(-1)^{\l(r(w))}=(-1)^{\frac{j-i-|w_0|+1}{2}}$.

\item[(iii)] Passing from summation over all $w\in W(\tau)$ to summation over all $w\in W_{kl}$, for $k=1,2$, $l=3,4$, we divide by $2^{\delta_{i0}+\delta_{j0}}$ since the affected subtrees (with $i=0$ or $j=0$) are counted multiple.

\end{enumerate}

Finally, we obtain
\begin{equation*}
C(\tau)=\frac{3\alpha(\tau_{w_0})}{4|\Aut(\tau)|}\sum\limits_{k=1,2,\,l=3,4}
\sum\limits_{w\in W_{kl}}(-1)^j2^{-(\delta_{ia_k}+\delta_{ja_l}+\delta_{i0}+\delta_{j0})}.
\end{equation*}
Again, the claim follows by lemma \ref{lemma:rectangle}. This finishes the proof.
\end{proof}

\begin{rk}
Note that while in the situation of case (1) and (3) the factor $2/3$ in the definition of the functions $H_{2k}$, for  $k\in\N$, is not needed, in the situation of case (2) it is essential for the vanishing of $C(\tau)$.
\end{rk}

\begin{ex}
We illustrate equation (\ref{formula:Cel_trees_proof}) in table (\ref{tab:ex1}--\ref{tab:ex3}).
\end{ex}

\begin{table}[ht]
\centering
\renewcommand{\arraystretch}{1.9}
\begin{tabular}{|c|c|c|c|c|c|}
$w$ & $\alpha(\tau_w)$ & $\sigma(\tau_w)$ & $c(w)$ & $\omega_{\tau}(w)$ & $(-1)^{\l(r(w))}$ \\ 
\hline 
\begin{tikzpicture}[scale=0.3]
	\tikzstyle{vertex}=[circle,fill,scale=0.3]
	\node[vertex] (a1_1) at (-2,0) {};
	\node[vertex, color=green] (a1_2) at (-4,0) {};
	\node[vertex] (a2_1) at (2,0) {};
	\node[vertex] (a2_2) at (4,0) {};
	\node[vertex] (v0) at (0,0) {};
    \draw (a1_1) -- (v0) (a1_2) -- (a1_1) (a2_1) -- (v0) (a2_2) -- (a2_1);
\end{tikzpicture} & $2$ & $1$ & $1$ & $1$ & $1$ \\ 
\begin{tikzpicture}[scale=0.3]
	\tikzstyle{vertex}=[circle,fill,scale=0.3]
	\node[vertex, color=green] (a1_1) at (-2,0) {};
	\node[vertex] (a1_2) at (-4,0) {};
	\node[vertex] (a2_1) at (2,0) {};
	\node[vertex] (a2_2) at (4,0) {};
	\node[vertex] (v0) at (0,0) {};
    \draw (a1_1) -- (v0) (a1_2) -- (a1_1) (a2_1) -- (v0) (a2_2) -- (a2_1);
\end{tikzpicture} & $4$ & $1$ & $1$ & $1$ & $-1$ \\ 
\begin{tikzpicture}[scale=0.3]
	\tikzstyle{vertex}=[circle,fill,scale=0.3]
	\node[vertex] (a1_1) at (-2,0) {};
	\node[vertex] (a1_2) at (-4,0) {};
	\node[vertex] (a2_1) at (2,0) {};
	\node[vertex] (a2_2) at (4,0) {};
	\node[vertex, color=green] (v0) at (0,0) {};
    \draw (a1_1) -- (v0) (a1_2) -- (a1_1) (a2_1) -- (v0) (a2_2) -- (a2_1);
\end{tikzpicture} & $4$ & $2$ & $1$ & $1$ & $1$ \\ 
\begin{tikzpicture}[scale=0.3]
	\tikzstyle{vertex}=[circle,fill,scale=0.3]
	\node[vertex] (a1_1) at (-2,0) {};
	\node[vertex] (a1_2) at (-4,0) {};
	\node[vertex] (a2_1) at (2,0) {};
	\node[vertex] (a2_2) at (4,0) {};
	\node[vertex] (v0) at (0,0) {};
	\draw [color=green,line width=2pt] (a1_1) -- (v0) (a2_1) -- (v0);
    \draw (a1_1) -- (v0) (a1_2) -- (a1_1) (a2_1) -- (v0) (a2_2) -- (a2_1);
\end{tikzpicture} & $4$ & $2$ & $-\frac23$ & $2$ & $1$ \\ 
\begin{tikzpicture}[scale=0.3]
	\tikzstyle{vertex}=[circle,fill,scale=0.3]
	\node[vertex] (a1_1) at (-2,0) {};
	\node[vertex] (a1_2) at (-4,0) {};
	\node[vertex] (a2_1) at (2,0) {};
	\node[vertex] (a2_2) at (4,0) {};
	\node[vertex] (v0) at (0,0) {};
	\draw [color=green,line width=2pt] (a1_1) -- (v0) (a1_2) -- (a1_1);
    \draw (a1_1) -- (v0) (a1_2) -- (a1_1) (a2_1) -- (v0) (a2_2) -- (a2_1);
\end{tikzpicture} & $2$ & $1$ & $-\frac23$ & $2$ & $-1$ \\ 
\end{tabular} 
\caption{Case (1): $\equivw\in\W(\tau)$ and coefficients.}
\label{tab:ex1}
\end{table}

\begin{table}[ht]
\centering
\renewcommand{\arraystretch}{1.9}
\begin{tabular}{|c|c|c|c|c|c|}
$w$ & $\alpha(\tau_w)$ & $\sigma(\tau_w)$ & $c(w)$ & $\omega_{\tau}(w)$ & $(-1)^{\l(r(w))}$ \\ 
\hline 
\begin{tikzpicture}[scale=0.3]
	\tikzstyle{vertex}=[circle,fill,scale=0.3]
	\node[vertex] (a1) at (-2,0) {};
	\node[vertex] (a2) at (0,2) {};
	\node[vertex] (a3_1) at (2,0) {};
	\node[vertex] (a3_2) at (4,0) {};
	\node[vertex,color=green] (v0) at (0,0) {};
    \draw (a1) -- (v0) (a2) -- (v0) (a3_1) -- (v0) (a3_1) -- (a3_2);
\end{tikzpicture} & $8$ & $2$ & $1$ & $1$ & $1$ \\ 
\begin{tikzpicture}[scale=0.3]
	\tikzstyle{vertex}=[circle,fill,scale=0.3]
	\node[vertex] (a1) at (-2,0) {};
	\node[vertex] (a2) at (0,2) {};
	\node[vertex] (a3_1) at (2,0) {};
	\node[vertex] (a3_2) at (4,0) {};
	\node[vertex] (v0) at (0,0) {};
	\draw [color=green,line width=2pt] (a1) -- (v0) (a2) -- (v0);
    \draw (a1) -- (v0) (a2) -- (v0) (a3_1) -- (v0) (a3_1) -- (a3_2);
\end{tikzpicture} & $2$ & $1$ & $-\frac23$ & $1$ & $1$ \\ 
\begin{tikzpicture}[scale=0.3]
	\tikzstyle{vertex}=[circle,fill,scale=0.3]
	\node[vertex] (a1) at (-2,0) {};
	\node[vertex] (a2) at (0,2) {};
	\node[vertex] (a3_1) at (2,0) {};
	\node[vertex] (a3_2) at (4,0) {};
	\node[vertex] (v0) at (0,0) {};
	\draw [color=green,line width=2pt] (a1) -- (v0) (a3_1) -- (v0);
    \draw (a1) -- (v0) (a2) -- (v0) (a3_1) -- (v0) (a3_1) -- (a3_2);
\end{tikzpicture} & $4$ & $2$ & $-\frac23$ & $4$ & $1$ \\ 
\begin{tikzpicture}[scale=0.3]
	\tikzstyle{vertex}=[circle,fill,scale=0.3]
	\node[vertex] (a1) at (-2,0) {};
	\node[vertex] (a2) at (0,2) {};
	\node[vertex] (a3_1) at (2,0) {};
	\node[vertex] (a3_2) at (4,0) {};
	\node[vertex] (v0) at (0,0) {};
	\draw [color=green,line width=2pt] (a3_1) -- (v0) (a3_2) -- (a3_1);
    \draw (a1) -- (v0) (a2) -- (v0) (a3_1) -- (v0) (a3_1) -- (a3_2);
\end{tikzpicture} & $4$ & $2$ & $-\frac23$ & $2$ & $-1$ \\ 
\end{tabular} 
\caption{Case (2):  $\equivw\in\W(\tau)$ and coefficients.}
\label{tab:ex2}
\end{table}

\begin{table}[ht]
\centering
\renewcommand{\arraystretch}{1.9}
\begin{tabular}{|c|c|c|c|c|c|}
$w$ & $\alpha(\tau_w)$ & $\sigma(\tau_w)$ & $c(w)$ & $\omega_{\tau}(w)$ & $(-1)^{\l(r(w))}$ \\ 
\hline 
\begin{tikzpicture}[scale=0.3]
	\tikzstyle{vertex}=[circle,fill,scale=0.3]
    \node[vertex] (vs) at (-2,0) {};
	\node[vertex] (vt) at (2,0) {};
	\node[vertex] (a1) at (-4,0) {};
	\node[vertex] (a2) at (-2,2) {};
    \node[vertex] (b1) at (4,0) {};
	\node[vertex] (b2_1) at (2,2) {};
	\node[vertex] (b2_2) at (2,4) {};
	\node[vertex] (v0) at (0,0) {};
	\draw [color=green,line width=2pt] (vs) -- (v0) (v0) -- (vt);
    \draw (vs) -- (vt) (vs) -- (a1) (vs) -- (a2) (vt) -- (b1) (vt) -- (b2_1) (b2_1) -- (b2_2);
\end{tikzpicture} & $16$ & $6$ & $-\frac23$ & $6$ & $1$ \\ 
\begin{tikzpicture}[scale=0.3]
	\tikzstyle{vertex}=[circle,fill,scale=0.3]
    \node[vertex] (vs) at (-2,0) {};
	\node[vertex] (vt) at (2,0) {};
	\node[vertex] (a1) at (-4,0) {};
	\node[vertex] (a2) at (-2,2) {};
    \node[vertex] (b1) at (4,0) {};
	\node[vertex] (b2_1) at (2,2) {};
	\node[vertex] (b2_2) at (2,4) {};
	\node[vertex] (v0) at (0,0) {};
	\draw [color=green,line width=2pt] (a1) -- (vs) (vs) -- (v0) (v0) -- (vt) (vt) -- (b1);
	\draw (vs) -- (vt) (vs) -- (a1) (vs) -- (a2) (vt) -- (b1) (vt) -- (b2_1) (b2_1) -- (b2_2);
\end{tikzpicture} & $4$ & $1$ & $\frac23$ & $2$ & $1$ \\ 
\begin{tikzpicture}[scale=0.3]
	\tikzstyle{vertex}=[circle,fill,scale=0.3]
    \node[vertex] (vs) at (-2,0) {};
	\node[vertex] (vt) at (2,0) {};
	\node[vertex] (a1) at (-4,0) {};
	\node[vertex] (a2) at (-2,2) {};
    \node[vertex] (b1) at (4,0) {};
	\node[vertex] (b2_1) at (2,2) {};
	\node[vertex] (b2_2) at (2,4) {};
	\node[vertex] (v0) at (0,0) {};
	\draw [color=green,line width=2pt] (a1) -- (vs) (vs) -- (v0) (v0) -- (vt) (vt) -- (b2_1);
	\draw (vs) -- (vt) (vs) -- (a1) (vs) -- (a2) (vt) -- (b1) (vt) -- (b2_1) (b2_1) -- (b2_2);
\end{tikzpicture} & $8$ & $6$ & $\frac23$ & $12$ & $1$ \\ 
\begin{tikzpicture}[scale=0.3]
	\tikzstyle{vertex}=[circle,fill,scale=0.3]
    \node[vertex] (vs) at (-2,0) {};
	\node[vertex] (vt) at (2,0) {};
	\node[vertex] (a1) at (-4,0) {};
	\node[vertex] (a2) at (-2,2) {};
    \node[vertex] (b1) at (4,0) {};
	\node[vertex] (b2_1) at (2,2) {};
	\node[vertex] (b2_2) at (2,4) {};
	\node[vertex] (v0) at (0,0) {};
	\draw [color=green,line width=2pt] (vs) -- (v0) (v0) -- (vt) (vt) -- (b2_1) (b2_1) -- (b2_2);
	\draw (vs) -- (vt) (vs) -- (a1) (vs) -- (a2) (vt) -- (b1) (vt) -- (b2_1) (b2_1) -- (b2_2);
\end{tikzpicture} & $8$ & $6$ & $\frac23$ & $6$ & $-1$ \\ 
\end{tabular} 
\caption{Case (3): $\equivw\in\W(\tau)$ and coefficients.}
\label{tab:ex3}
\end{table}


\section{Conclusions}
In this paper, we have considered an Ansatz for the study of the existence of formal integrals for Kahan-Hirota-Kimura discretizations.
We have used the formalism of trees (related to B-series) to develop a combinatorial proof of formula (\ref{CMOQ}) for a modified integral of the Kahan-Hirota-Kimura discretization in the case of Hamiltonian systems on symplectic vector spaces or Poisson vector spaces with constant Poisson structures. The proof presented in this work gives insights into the combinatorial structure that ensures the existence of a modified integral in this case.

Goals for future research are to deepen the study of the partial differential equations (\ref{eqnEpsilon}) in order to further the understanding of the mechanism that ensures (or prevents) that the Kahan-Hirota-Kimura discretization admits integrals of motion.

\section*{Acknowledgements}
The author would like to thank Matteo Petrera and Yuri Suris for inspiring discussions and their critical feedback on this manuscript.

\end{document}